# Cascades of Fano Resonances in Scattering by a Mesoscale Spherical Particle in the Superresonance Mode

Igor V. Minin[a], Song Zhou[b], and Oleg V. Minin[a]
[a]Tomsk Polytechnic University, 30 Lenin Ave., Tomsk 634050 Russia
[b]Jiangsu Key Laboratory of Advanced Manufacturing Technology, Faculty of Mechanical and Material Engineering, Huaiyin Institute of Technology, Huai'an 223003, China.

**Abstract**
Broadband light illumination of a mesoscale dielectric sphere makes it possible to reveal new effects that associated with super resonance mode. These include the possibility of generating high-order Fano resonance cascades. The quality factor has the order $Q=10^7$. Super-resonance-enabled subdiffraction fields localization by mesoscale dielectric sphere under broadband light illuminations (e.g., at wavebands of 400…700 nm for refractive index of sphere n=1.5-1.9 and 1500…1600 nm for n=3.47) have been investigated. The conditions for the quasi-periodicity of superresonance peaks are established. The dependence of the amplitude modulation of Fano cascades on the refractive index of the sphere is shown. These results are important in deep understanding of physics of the super-resolution mechanism related to superresonance mode and will find great potential applications in many other area.
**Keywords:** *superresonance, Fano cascades, high-order Mie resonance,*

**Introduction**

Fano resonances are among the well-known optical phenomena of narrow and broad spectrum interference [1–3]. Due to interference, the symmetric Lorentzian contour of a narrow line is transformed into an asymmetric contour [4-5], the shape of which is described in terms of the Fano theory [1-8]. Usually, a pronounced Fano resonance in dielectric spheres is observed under the condition that the characteristic Mie size parameter $q$ is about unity, $q=2\pi a/\lambda \approx 1$, where $a$ is the radius of a spherical particle in our case, $\lambda$ is the radiation wavelength [6–9].

Previously, Fano interference involving nonresonant Mie scattering was observed in photonic crystals [10, 11], in interference between different Mie modes [9], and also between nonresonant and resonant scattering on spherical particles with a diameter less than a wavelength [12, 13]. The interference between a plasmon dipole mode and various controlled modes in a water guide, leading to three sharp Fano resonances, was considered in [14]. Fano resonance in the scattering of an electromagnetic wave by a dielectric sphere with a size parameter of about 1 and a high refractive index of about 30 was analyzed in [15]. The optical properties of a microresonator whose transmission spectrum essentially consisted of a single mode were studied in [16].

Features of the Fano resonance in dielectric spheres with a diameter near the wavelength and a refractive index of the material less than 2 were considered in [17]. Scattering by nonmagnetic spherical particles from a material with a permittivity of less than 2 and a size parameter of less than 10 was studied in [18]. It was shown that in this case closely packed multipole resonances were found, in which both magnetic and electric multipolar moments have complex Mie amplitudes ($b_n$ and $a_n$) with comparable amplitudes.

Fano resonances during the diffraction of axially symmetric Bessel beams on a dielectric sphere with a refractive index from 1.64 to 2 and a size parameter less than 10 were discussed in [19]. Fano resonances were also observed in a terahertz spherical resonator with a whispering gallery mode coupled to a multimode waveguide [20].

At the same time, Fano interference effects are observed in mesoscale structures, which have been little studied today, with a Mie size parameter of the order of $q \sim 10$. For example, Fano-type resonances have been observed in photonic molecules consisting of several spheres with a size parameter $q \approx 20$ and a refractive index of about 1.8 [21]. We note that one of the unusual properties of weakly dissipative dielectric mesoscale structures [22–24] is associated with the simultaneous coexistence of two effects, topological optics [25] and the Fano resonance, at which the characteristic size of optical vortices inside a particle is much smaller than the diffraction limit [22–30]. In these works, a relationship between Fano resonances in light scattering by mesodimensional dielectric structures and subwavelength optical vortices was revealed. Such optical vortices arise under the condition that the dielectric particle size parameter is greater than a certain value, which depends on its optical contrast [28].

In particular, the effect of superresolution in a mesoscale diffractive structure based on anomalous apodization effect is due to the specific arrangement of optical vortices inside the particle and Fano resonances excited in the phase array due to the effective coupling of controlled Fabry-Perot resonances and Mie-like structural resonances [26]. It was found in [27, 29] that mesoscale spheres can stimulate extremely high field enhancement at singularities, which then form two circular hotspots around the sphere poles. At the same time, the existence of an unusual large three-dimensional circulation of the Poynting vector in the form of a "heart" was discovered for the first time [29]. For spherical particles with $q\sim 10$, at the conditions of Fano resonances of high orders, the observed degree of space localization of both magnetic and electric fields is far beyond the diffraction limit, both inside the particle and near on its surface [23-32]. In this case, it is surprising that the introduction of a small dissipation into the material of a spherical particle can enhance the spatial localization of the field [30].

Previously, based on the Lorentz-Mie theory, it was shown that as a result of the scattering of a plane wave on circular infinite rods with a size parameter less than 5 and with a high refractive index of the dielectric ($\varepsilon=10\ldots 50$), Fano resonance cascades can occur [33]. However, these results cannot be directly applied to the conditions of superresonance [27, 34], which is supported only by spherical particles [23–32, 34, 35]. In [27], the intensity spectrum of electric and magnetic fields and the high enhancement of these fields associated with them were presented on the shadow surface of a dielectric sphere with $q=19-21$ and a refractive index n=1.5, containing 4 Fano resonance lines. It is noteworthy that these resonances are present for all multipolar orders, and their intensity increases linearly with the multipolar order [27]. We note that with an increase in the refractive index of the sphere material, the number of the resonant mode decreases and the maximum achievable field intensity under superresonance conditions increases. However, for a refractive index greater than two, the positions of the field localization shifts in the direction from the outer boundary of the sphere to its center, which can be qualitatively explained based on the formula for the focus of a spherical lens. In turn, an increase in the size of the spheres also leads to the excitation of more pronounced superresonance modes with stronger field localization. It should also be noted that these high-order internal Mie modes differ from other types of resonances in the dielectric sphere [36-40].

Unfortunately, high-order Fano resonance cascades in mesoscale ($q\sim 20\ldots 80$) dielectric spheres with a refractive index of about 1.5 or more still remain practically unstudied, despite their attractive characteristics. What little has been done so far indicates that it is possible to generate a high-order Fano resonance cascade with extremely narrow lines and gigantic intensity in the superresonance mode, thereby providing additional information about the properties of spherical particles and opening up new possibilities in various fields of science and technology. Therefore, a comprehensive study of high-order Fano resonance cascades in mesoscale dielectric spherical particles is necessary. Such an analysis in the literature has not been studied thus far.

**Main part**

Previously, we showed that non-dissipative mesoscale spherical particle, immersed in vacuum, could support high-order internal Mie modes [27, 29-32] and these high order Fano resonances we define as "superresonances" [27]. The nature of superresonance is that a single term with a high-order resonant mode can lead to a multiple increase in the scattered electric and magnetic fields intensity under certain conditions [27,31,32]. For example, on the spectrum of magnetic and electric field intensities for the dielectric sphere with $n=1.9$ and $q\sim 31\ldots 33$ immersed in water [31] (optical contrast of the spherical particle is 1.43), one can notice the alternation of the prevailing

14 intensities peaks (Figure 1). We also note that under the conditions under consideration (in a given range of size parameters of the sphere and its optical contrast), the maximum amplitude of both magnetic and electrical superresonances changes in waves.

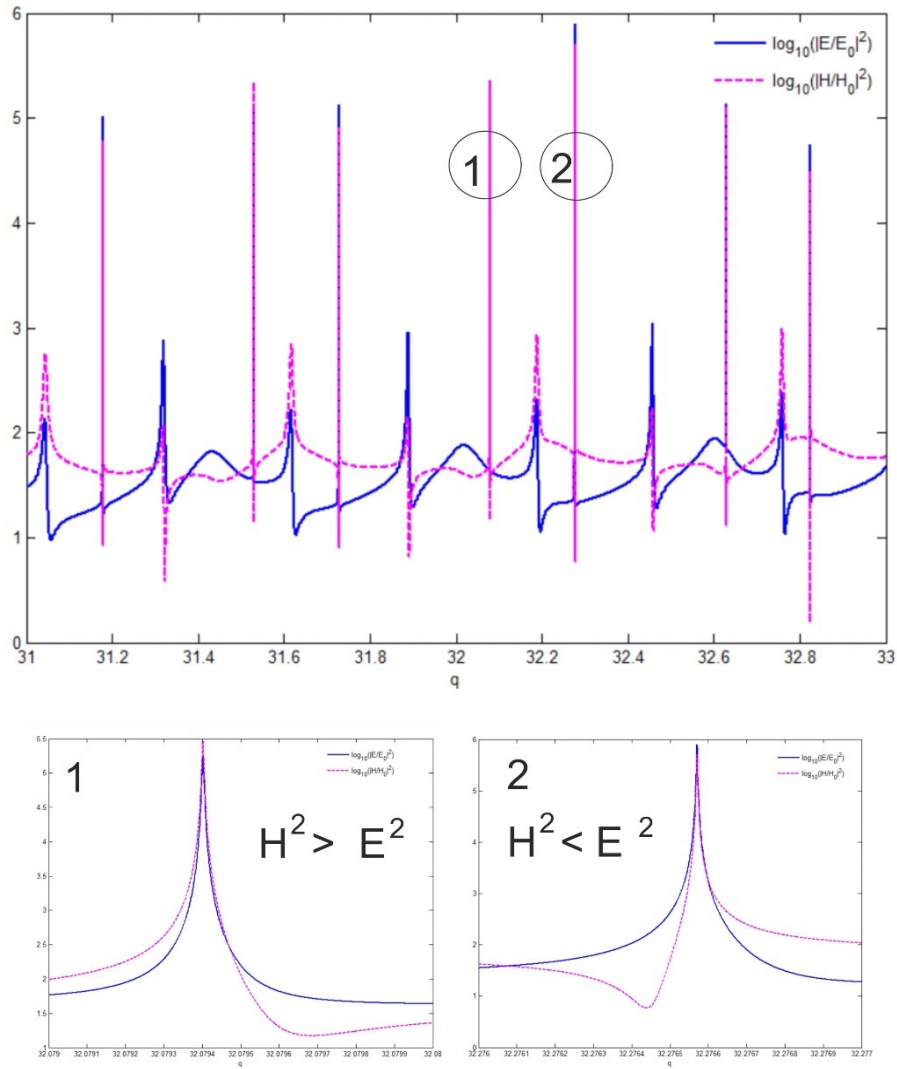

Figure 1. Intensities of magnetic and electric fields with a profile characteristic of a pronounced cascade of Fano resonances on the shadow surface of a particle with a refractive index of $n=1.9$ and a size parameter of about $q \approx 30$ located in water (optical contrast of the sphere 1.43) [31]. One can see the alternation of predominant magnetic and electric resonances. Below are shown 2 resonant intensities in the range of the size parameters $q=32.07941$ (left) and $q=32.27657$ (right) of the same particle, characterized by an extremely narrow width of the resonance line.

As can be seen from Fig. 1, the observed spectrum demonstrates a strongly asymmetric nature of the observed superresonances, which have the shapes of Fano lines [2,5,6]. In this case, all resonance lines are very sharp. One of the determining factors is the spectral narrowness of the particle's own excitations and, accordingly, the narrowness of the emitted resonance bands. According to the definition, the quality factor Q of resonator is $Q=f_r/\Delta f = q/\Delta q$ [41,42], where $f_r$ is the resonant frequency, $\Delta f$ is the resonance width (full width at half maximum [43]). It follows from Figure 1 that the quality factor Q has the order $Q=5.8 \cdot 10^6$ for $q=32.0794016$ with the width of the resonance line $\Delta q=5.52 \cdot 10^{-6}$ and $Q=1.2 \cdot 10^7$ for $q=32.2765704$ with half the width of the resonance line $\Delta q= 2.73 \cdot 10^{-6}$. The linewidths for E and H intensities and, accordingly, the quality factor are approximately the same.

The decisive role in the occurrence of Fano resonances is played by magnetic dipole resonances of isolated dielectric particles. The magnetic dipole mode of a dielectric particle excited at the magnetic resonance wavelength can be stronger than the response of the electric dipole, and thus make the main contribution to the scattering efficiency (see Fig. 2). Such spherical particles have a unique location of hot spots at the poles of the sphere [30, 37], due to the specific behavior of the internal Mie modes, with a high degree of spatial localization of magnetic and electric fields exceeding the diffraction limit, both inside the particle and on its surface [27, 29-32, 35]. Figure 1 also clearly shows that the asymmetry of the resonance intensities is mirrored when the magnetic or electric field prevails, and the width of the resonance line is greater for the field, which corresponds to the intensity maximum. Moreover, the cascades of Fano resonances observed in the scattering of radiation by weakly dissipative dielectric mesoscale spheres are accompanied by singular phase effects with large values of the local wave vector [24] and the generation of optical vortices with a characteristic spatial size much smaller than the diffraction limit [16, 18, 19, 27,34,35,37,38].

The intensity of the peaks in the spectrum of resonant characteristics (Fig. 1) alternate - the peak corresponding to superresonance and the peak corresponding to the "ordinary" resonance with relative small intensity. At the same time, their intensity differs by more than 2 orders of magnitude, and their extrema are in antiphase. It is also interesting to note that in the cascade of superresonance modes (Fig. 1) there are pronounced characteristic alternating subcascades with periods of about $\Delta q \approx 0.35$ and $\Delta q \approx 0.20$, shown in Figure 2, but the absolute values of the intensities of magnetic and electric fields differ depending on the type mode (magnetic or electric) and size parameter. This quasi-periodicity can find a number of interesting applications, including the selection of spheres to achieve the best field localization, improve the resolution of subwavelength focusing, and possibly to create a quasi-periodic frequency comb.

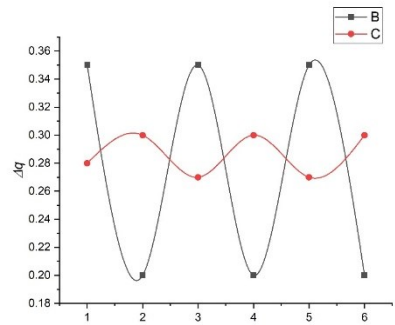

Figure 2. Periodicity of resonant modes upon scattering by a dielectric sphere in water. The black curve corresponds to superresonant modes (B), the red curve corresponds to "ordinary" resonances with an intensity more than 2 orders of magnitude lower (C).

Let us now briefly consider the spectral characteristics of superresonance cascades upon irradiation of a spherical particle with fixed values of the refractive index and diameter. At the same time, without reducing the generality of the problem, to simplify the simulations and taking into account the results of [31,32,35], we will assume that the sphere is located in a vacuum.

The spectrum of superresonances for a spherical particle with a refractive index of 1.5 and a radius a = 2.5 µm when it is irradiated with radiation in the wavelength range from 400 nm to 700 nm is shown in Figure 3. In particular, Figure 3 shows that resonance can lead to the complete disappearance of scattering at a certain wavelength (frequency) or along a certain direction.

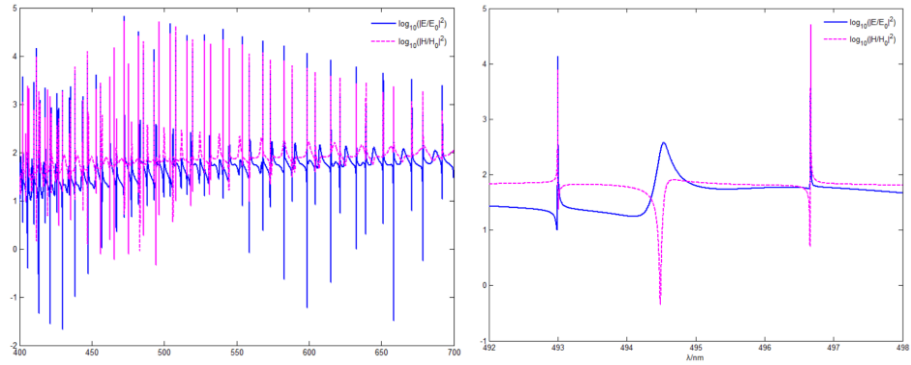

Figure 3. Left: Fields intensities at the shadow pole of sphere in vacuum, n=1.5, a =2.5 mkm, , wavelength = 400…700 nm (q=22…39) with step 0.001 nm. Right: Fano-like lines for 2 resonant wavelengths 493.002 nm and 496.668 nm.

The change in the spectral characteristics of superresonance when the observation point is removed from the shadow surface of the sphere pole by 100 nm is shown in Figure 4.

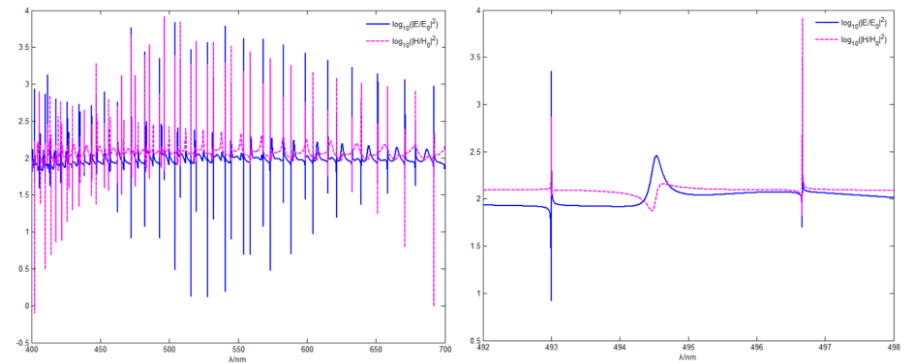

Figure 4. The same as in Fig.3 at the point on 100 nm from the sphere surface.

The alternation of electric and magnetic modes in superresonant lines is explained by the characteristic behavior of the modes, which in turn are characterized by the corresponding Mie scattering coefficients (details can be found in [31]). For some resonant wavelengths, the corresponding values are given in Table 1.

Table. Resonant modes vs mail resonant wavelength.

| $\lambda$ | $A_{l=}^{m}$ | $A_{l=}^{e}$ |
|---|---|---|
| 462.522 | 45 | |
| 465.962 | | 44 |
| 504.092 | 41 | |
| 507.950 | | 40 |
| 554.056 | 37 | |
| 558.859 | | 36 |

It can be seen from the above results that as the wavelength increases, the number of the resonant mode decreases, and the resonances themselves alternate (magnetic-electric). In this case, the spectral position of the resonant peaks can be controlled by changing the effective size parameter (in this case, the wavelength of the incident radiation) and the relative refractive index of the sphere material, as well as the conditions of the external environment.

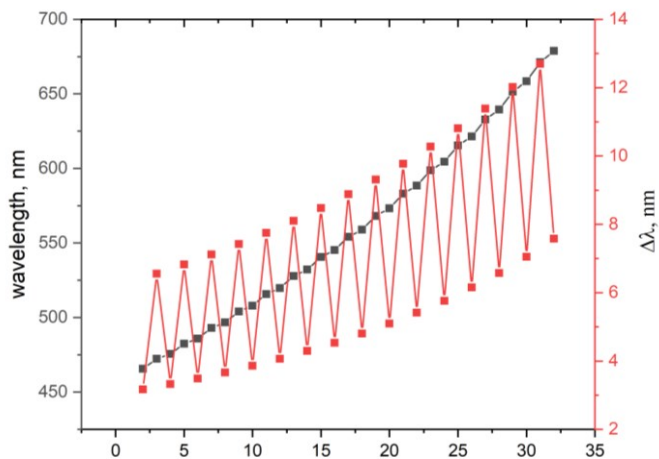

Figure 5. Resonance wavelengths in the range of 400…700 nm and distances between resonant peaks for sphere in vacuum, n=1.5, a =2.5 mkm, corresponding to the spectrum shown in Figure 3. The values of resonant wavelengths are given below in Table 2.

Table 2. Resonant values of wavelengths.

| 462.522 | 465.692 | 472.247 | 475.571 |
|---|---|---|---|
| 482.397 | 485.886 | 493.002 | 496.668 |
| 504.092 | 507.950 | 515.702 | 519.768 |
| 527.871 | 532.163 | 540.641 | 545.177 |
| 554.056 | 558.859 | 568.170 | 573.263 |
| 583.037 | 588.450 | 598.721 | 604.485 |
| 615.294 | 621.445 | 632.832 | 639.410 |
| 651.426 | 658.479 | 671.175 | 678.757 |

It follows from Fig. 5 that with an increase in the wavelength irradiating the sphere, the resonant wavelength also slightly nonlinearly increases, while the period between resonant modes in the frequency domain also increases. Thus, at a resonant wavelength of about 462 nm, the distance to the neighboring peak of the superresonance mode is about 3.17 nm, and at a wavelength of 678 nm, it is 7.58 nm. For "neighboring" resonant wavelengths of 472 nm and 671 nm, the distances to neighboring peaks are 6.55 nm and 12.69 nm, respectively. That is, the periods of the spectral position of the superresonance modes change approximately 2 times when the wavelength changes from 460 nm to 670 nm.

Figures 6-7 show similar spectral characteristics for a sphere of the same radius, but with an increased refractive index to 1.9. It can be seen that with an increase in the refractive index of the material of a spherical particle from 1.5 to 1.9, the "frequency" of external modulation of the resonant peaks increases. In this case, the intensity of superresonant modes increases by about an order of magnitude.

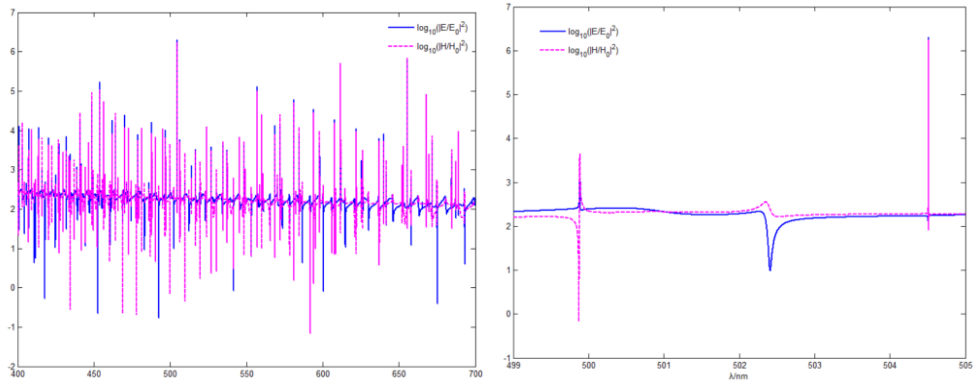

Figure 6. Fields intensities at the shadow pole of sphere in vacuum, n=1.9, a =2.5 mkm, , wavelength = 400…700 nm (q=22…39) with step 0.001 nm. Right: Fano-like lines for 2 resonant wavelengths.

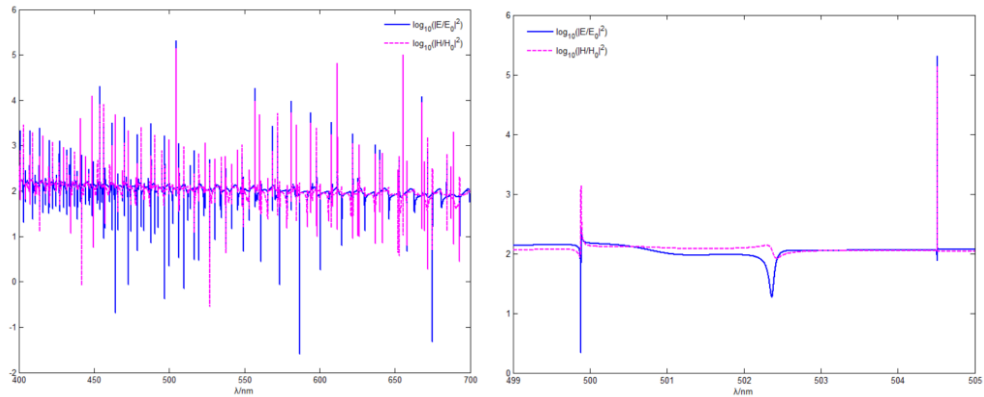

Figure 7. The same as in Fig.6 but at the point on 100 nm from the sphere shadow surface.

Similar dependences for a spherical particle with a refractive index of 3.47 in the telecommunications wavelength range from 1500 nm to 1600 nm are shown in Figures 8-9. The distribution of resonant mode intensities on the shadow surface of a spherical particle has a rather "chaotic" form, which is due to the fact that with a refractive index of more than 2, the field localization region is not near the poles of the sphere, but inside it. This limits a number of practical applications of such particles. This is illustrated in Figure 10. Note that the intensity of the magnetic field is approximately an order of magnitude higher than the electric one.

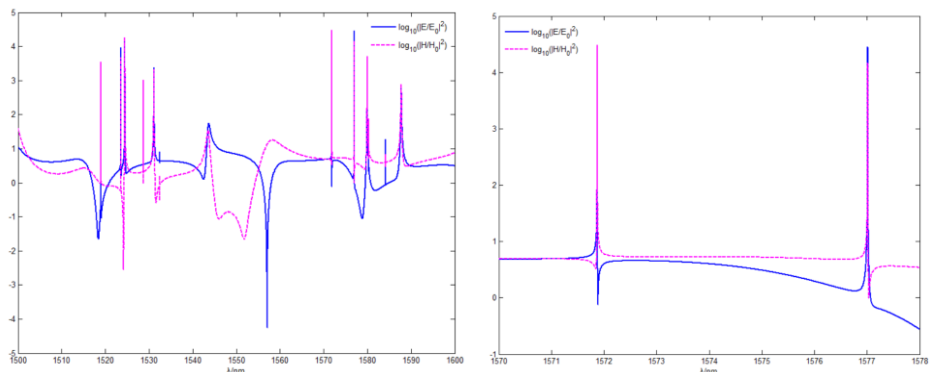

Figure 8. Fields intensities at the shadow pole of sphere in vacuum, n=3.47, a =2.5 mkm, at telecommunication wavelength = 1500…1600 nm with step 0.001 nm. Right: Fano-like lines for 2 resonant wavelengths.

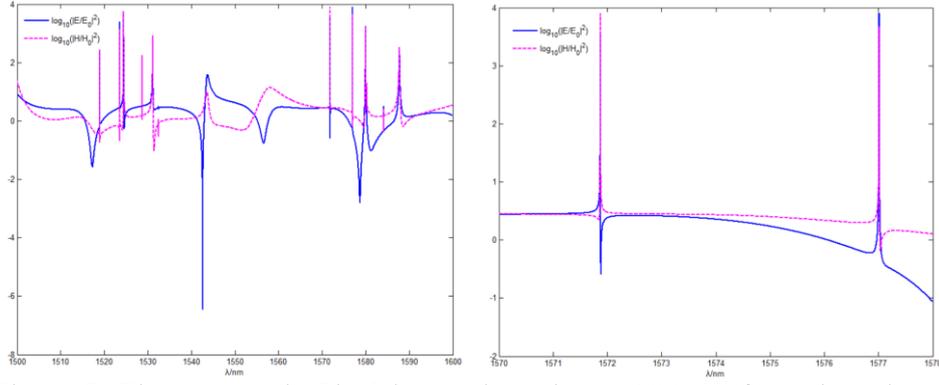

Figure 9. The same as in Fig.8 but at the point on 100 nm from the sphere surface.

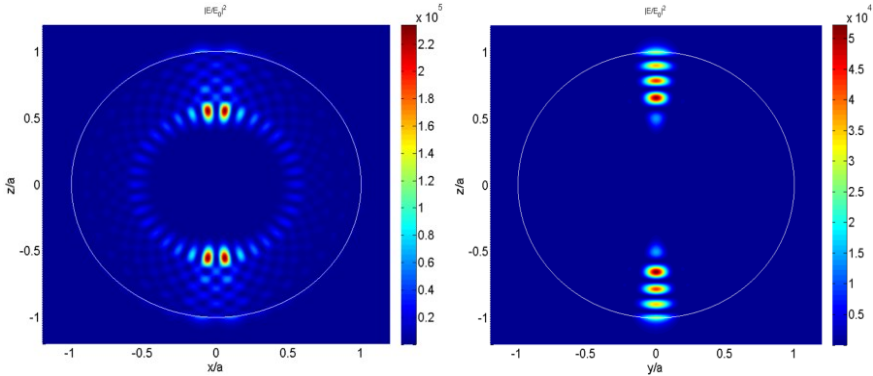

Figure 10a. Electric field intensity distributions for sphere with n=3.47 in 2 planes.

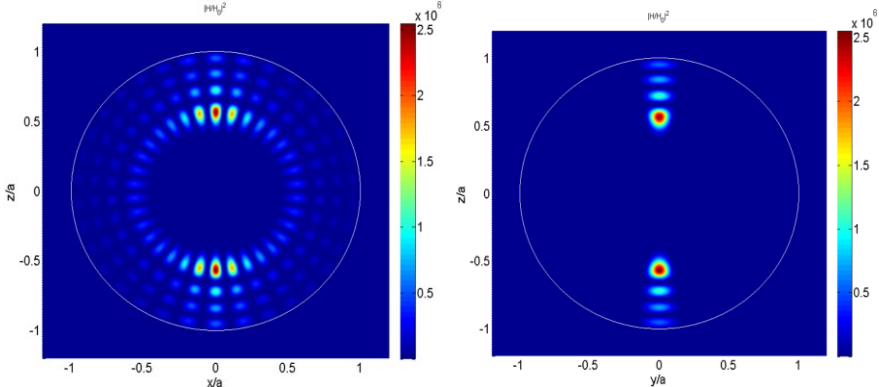

Figure 10b. Magnetic field intensity distributions for sphere with n=3.47 in 2 planes.

In next step works, we will deep investigated material loss [27,30] and dispersion in mesosphere materials to the focusing properties of spherical particle. To this end, in optics BK7 optical glass is preferable (refractive index near 1.5 and extinction coefficient k~$9.7 \cdot 10^{-9}$) at waveband 0.5 – 1.0 mkm.

**Conclusion**

The generation of extremely strong and limited fields on a deep subwavelength scale (far beyond the diffraction limit) plays a central role in modern plasmonics, magnonics, and in photonic and metamaterial based structures. Due to their unusual and unique physical properties, high-order Fano resonances have a good potential for application in various fields, from the generation of extremely high fields [27] to ultrasensitive sensors [31,32,35], spectral filters and switches. Moreover, their intriguing properties have made them part and parcel of cutting-edge research in the optical and microwave (THz) regime.

In the present work, within the framework of the rigorous Lorentz-Mie theory, it is shown that mesoscale low loss dielectric spheres can be used to generate high-order Fano resonance cascades with giant magnetic and electric field intensities, and their main properties are considered. We believe that our research can significantly expand the understanding of the seemingly well-known resonant Mie scattering and may open up new possibilities for manipulating and controlling electromagnetic waves with extreme intensities using mesoscale dielectric spheres with both low and high refractive indices. Mie scattering in the superresonance mode in space near the boundary of a scattering dielectric sphere (size parameter $q$~10…50, refractive index 1.5…4) is a cascade of Fano resonances. It is shown that due to the interference nature of the resonant Mie scattering, the superresonance mode manifests itself in the form of cascades of asymmetric resonant lines, the shape of which is described in terms of the Fano theory. These results are important in deep understanding the super-resolution mechanism related to superresonance mode and some of the important applications in different physical area.

**Acknowledgements**
This work was carried out within the framework by the Tomsk Polytechnic University Development Program, FRBR grant (21-57-10001) and Natural Science Research Program of Huai'an (No. HAB202153).